\documentclass[disable]{article} %add [disable] to remove notes

\usepackage[linesnumbered,ruled,vlined]{algorithm2e}
\usepackage{amssymb}
\usepackage{amsmath}
\numberwithin{figure}{section} 
\numberwithin{table}{section} 
\usepackage{array}
\usepackage{authblk}
\usepackage{siunitx}

\usepackage[sorting=none,citestyle=numeric-comp]{biblatex} 
\usepackage{booktabs} 

\usepackage{cjhebrew}
\usepackage{changepage}
\usepackage{changes}
\usepackage{chemformula} 

\usepackage{enumitem}
\usepackage{floatpag}
\usepackage{footnote}

\usepackage[letterpaper,top=2cm,bottom=2cm,left=3cm,right=3cm,marginparwidth=2cm]{geometry}
\usepackage{graphicx}

\usepackage[T1]{fontenc}

\usepackage{hyperref} 
\hypersetup{colorlinks=true, allcolors=blue}
\usepackage{cleveref} 

\usepackage{lscape}
\usepackage{mathtools}
\usepackage[version=4,arrows=pgf-filled, textfontname=sffamily, mathfontname=mathsf]{mhchem}
\usepackage{multirow}

\usepackage{physics}
\usepackage{subcaption}
\usepackage{tablefootnote}
\usepackage{titlesec} 
\usepackage[normalem]{ulem} 
\usepackage{textcomp}

\usepackage{colortbl}

\usepackage{amsthm}
\theoremstyle{definition}

\theoremstyle{remark}

\newcommand\note[1]{\todo[inline,color=teal!30]{#1}}

\usepackage{setspace}

\title{QB Ground State Energy Estimation Benchmark}

\begin{document}
\author[1,2,3]{Nicole Bellonzi\thanks{Corresponding author. Email: bellonzi@apollo-quantum.com}}
\author[4,5]{Joshua T. Cantin}
\author[4,5]{Mohammad Reza Jangrouei}
\author[2,3]{Alexander Kunitsa}
\author[6]{Jason Necaise}
\author[7]{Nam Nguyen}
\author[2]{John Penuel}
\author[1,2,3]{Maxwell D. Radin}
\author[2,3]{Jhonathan Romero Fontalvo}
\author[8]{Rashmi Sundareswara}
\author[4,5]{Linjun Wang}
\author[8]{Thomas Watts}
\author[3]{Yanbing Zhou}
\author[2]{Michael C. Garrett}
\author[8]{Adam Holmes}
\author[4,5]{Artur F. Izmaylov}
\author[9]{Matthew Otten\thanks{Corresponding author. Email: mjotten@wisc.edu}}

\affil[1]{Apollo Quantum LLC, Cambridge, MA, USA}
\affil[2]{L3Harris Technologies, Inc., Palm Bay, FL, USA}
\affil[3]{Zapata AI Inc., Boston, MA 02110 USA}
\affil[4]{Chemical Physics Theory Group, Department of Chemistry, University of Toronto, Toronto, Ontario M5S 3H6, Canada}
\affil[5]{Department of Physical and Environmental Sciences, University of Toronto Scarborough, Toronto, Ontario M1C 1A4, Canada}
\affil[6]{Dartmouth College, Hanover, NH, USA}
\affil[7]{Applied Mathematics, Boeing Research \& Technology, Huntington Beach, CA, USA}
\affil[8]{HRL Laboratories, Malibu, CA, USA}
\affil[9]{University of Wisconsin -- Madison, Madison, WI, USA}

\maketitle

\begin{abstract}
Ground State Energy Estimation (GSEE) is a central problem in quantum chemistry and condensed matter physics, demanding efficient algorithms to solve complex electronic structure calculations. This work introduces a structured benchmarking framework for evaluating the performance of both classical and quantum solvers on diverse GSEE problem instances. We assess three prominent methods -- Semistochastic Heat-Bath Configuration Interaction (SHCI), Density Matrix Renormalization Group (DMRG), and Double-Factorized Quantum Phase Estimation (DF QPE) --highlighting their respective strengths and limitations. Our results show that fully optimized SHCI achieves near-universal solvability on the benchmark set, DMRG excels for low-entanglement systems, and DF QPE is currently constrained by hardware and algorithmic limitations.
However, we observe that many benchmark Hamiltonians are drawn from datasets tailored to SHCI and related approaches, introducing a bias that favors classical solvers. To mitigate this, we propose expanding the benchmark suite to include more challenging, strongly correlated systems to enable a  more balanced and forward-looking evaluation of solver capabilities. As quantum hardware and algorithms improve, this benchmarking framework will serve as a vital tool for tracking progress and identifying domains where quantum methods may surpass classical techniques.
The QB-GSEE benchmark repository is openly available at \url{https://github.com/isi-usc-edu/qb-gsee-benchmark}
\cite{qb-gsee-benchmark-repo}. By maintaining a scalable and open resource, we aim to accelerate innovation in computational quantum chemistry and quantum computing.

\end{abstract}

\pagebreak

\todo[inline,color=white]{To update Figures:
    (i) Download latest from the repo for any included figures. 
    (ii) Upload and replace in figures/ directory
    (iii) Update dates for the bib-tex citations. 
}

% \note{Introduction does not have a header. Shooting for 1, maybe 1.5, pages.}
\section{Introduction}
Ground State Energy Estimation (GSEE) has long served as a cornerstone problem in chemistry and physics, enabling precise calculations of chemical reaction rates and material properties. However, despite its centrality, GSEE also highlights fundamental computational challenges and limitations, particularly in balancing accuracy, efficiency, and scalability.

For decades, GSEE has driven advances in physics and computing, with algorithms evolving to address its complexities. Exact solutions remain exponentially hard to compute due to the rapid growth of the Hilbert space with system size, making them impractical for large systems. As a result, heuristic algorithms, ranging from approximations like Density Functional Theory (DFT) to methods like semistochastic heat-bath configuration interaction (SHCI), have become essential. These heuristics often provide useful and accurate results but operate with varying degrees of reliability, often lacking systematic guarantees of performance. The effectiveness of these approaches can sometimes be more art than science, relying on empirical tuning and problem-specific adaptations rather than universally applicable principles.

Despite these challenges, emerging technologies offer promising alternatives to classical heuristics. Quantum computing, in particular, represents a paradigm shift in computational science, with its potential for exponential advancements over the coming decade \cite{GoogleQuantumRoadmap, MicrosoftQuantumRoadmap, IBMQuantumRoadmap}. 
Among its many applications, quantum computing offers a fundamentally new approach to GSEE by naturally representing and manipulating quantum states, directly encoding wavefunctions within their qubits. 
Algorithms such as Variational Quantum Eigensolver (VQE) and Quantum Phase Estimation (QPE) leverage quantum principles to estimate ground-state energies more efficiently, offering the potential for polynomial or exponential speedups over classical methods. Additionally, quantum computers provide a pathway to solve problems with high accuracy, capturing all system correlations and surpassing the approximations required by classical methods like density functional theory.
However, the field remains in its early stages, with significant challenges in validating and benchmarking quantum algorithms for GSEE. Limited qubit coherence times, noise, and resource constraints hinder practical implementation, making it difficult to determine whether quantum algorithms truly outperform classical alternatives in meaningful scenarios. 

To effectively assess progress in GSEE and quantum computing, robust benchmarking frameworks are essential. Existing repositories such as HamLib \cite{Sawaya2024hamlib}, VarBench \cite{varbench}, and g2 \cite{g2neutral,g2ion} offer valuable datasets for evaluating both classical and quantum algorithms, yet they often lack comprehensive reference energies and detailed performance breakdowns, limiting their ability to explain when and why solvers succeed or fail. This makes it difficult to compare methods meaningfully across diverse Hamiltonian instances.
A number of studies have also defined benchmarks that seek to evaluate the performance of a quantum processor on tasks related to GSEE~\cite{Lubinski2023benchmarks,lubinski2024exploration,Chen2024benchmarking,AppQSim,Chatterjee2024benchmarking,supermarq,Li2023benchmark}.
While these benchmarks provide a way to compare present-day Noisy Intermediate Scale Quantum (NISQ) processors, it is challenging to relate the benchmark results to utility: in many cases the tasks being benchmarked represent only a subroutine within GSEE and, moreover, are derived from problem instances that are not representative of real-world use cases.
In addition, metrics for evaluating quantum computers can be misleading when the total time to solution is not accounted for or when circuit compilation tricks allow a solver to achieve a performance on benchmark tasks that is not representative of the performance on high-utility problem instances \cite{Smolin2013Oversimplifying,quantinuum_debunking_2024}.

\begin{figure}[tb]
    \centering
    \includegraphics[width=0.75\linewidth]{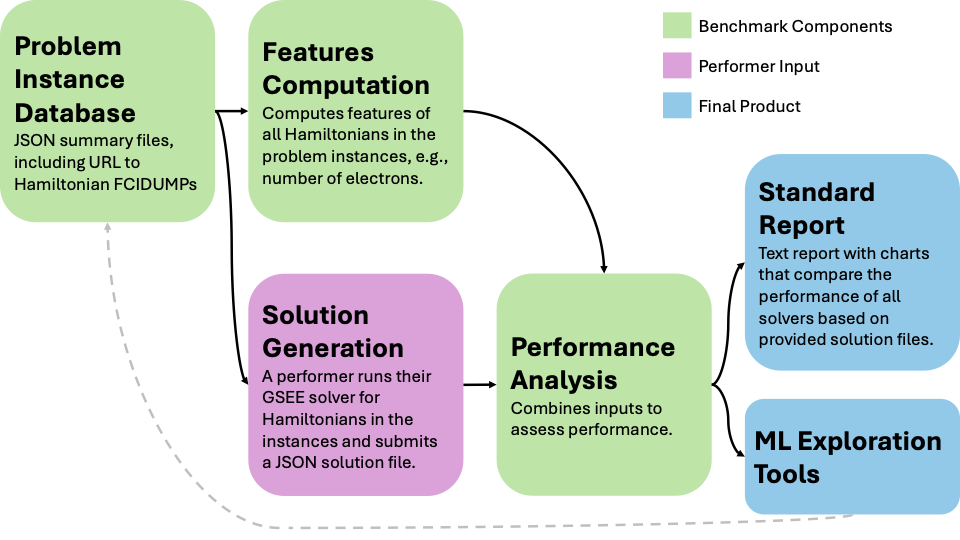}
    \caption{Overview of the main components of the GSEE benchmark repository \cite{qb-gsee-benchmark-repo}.}
    \label{fig:benchmarkframework}
\end{figure}

To help guide future advancements in both classical and quantum GSEE approaches, this work introduces an open-source GSEE benchmark that incorporates hardness metrics, systematically measuring algorithmic performance and identifying cases where GSEE algorithms excel or struggle.
By integrating diverse Hamiltonian problem sets, advanced feature extraction techniques, and machine learning-driven performance analyses, this benchmark can provide deeper insights into solver behavior, scalability, and generality.
This paper describes the benchmark and demonstrates its application to the Semistochastic Heat-Bath Configuration Interaction (SHCI), Density Matrix Renormalization Group (DMRG), and resource estimates for Double-Factorized Quantum Phase Estimation (DF QPE). This modular framework, shown in Figure~\ref{fig:benchmarkframework} and available on GitHub \cite{qb-gsee-benchmark-repo}, integrates Hamiltonian problem instances, evaluates their features, and applies machine learning techniques to analyze algorithm performance.

In Section~\ref{sec:results}, we will introduce the benchmarking framework and dive into the benchmark results, highlighting the performance of SHCI, DMRG, and DF QPE as key solvers. Section~\ref{sec:discussion} explores key insights from these results and outlines future directions to enhance benchmarking methodologies and expand the dataset. In Section~\ref{sec:component-features}, we break down the essential problem features that define computational challenges and solver capabilities. Section~\ref{sec:problem-selection} details the diverse set of problem instances used for evaluation, spanning real-world chemistry and physics applications. Finally, Section~\ref{sec:ML} introduces our machine learning-driven approach to predicting solvability, leveraging latent space analysis to map out the boundaries of what can—and cannot—be solved efficiently.

\section{Results}\label{sec:results}
The QB GSEE benchmark is designed to provide a rigorous and scalable evaluation of algorithms for ground-state energy estimation, integrating both classical and quantum approaches within a unified assessment framework. At its core, the benchmark consists of three interdependent components: a structured problem instance database, an automated feature computation module, and a comprehensive performance analysis pipeline. Each component is designed to facilitate a direct comparison of algorithmic performance while ensuring scalability, modularity, and extensibility.

The problem instance database serves as the foundation of the benchmark, housing a diverse collection of Hamiltonians that span different domains of computational chemistry and condensed matter physics. These problem instances, detailed in Section~\ref{sec:problem-selection}, are categorized into two primary types: benchmark instances, which correspond to well-characterized problems with reliable classical reference solutions\footnote{Reference solutions refer to FCI-level converged ground state energies. These can be obtained either from published converged energy values or by achieving agreement (within less than half of chemical accuracy) between fundamentally different GSEE algorithms, such as SHCI and DMRG.}, and guidestar instances, which are selected based on their scientific or industrial importance despite their intractability by classical methods. Each instance is stored in a standardized format that includes Hamiltonian specifications, metadata, and links to FCIDUMP files, ensuring consistency across different solvers. The schema for these instances is designed to support extensibility\cite{problem_schema}, allowing researchers to contribute new problems that capture emerging computational challenges.

To systematically evaluate algorithmic efficiency and scalability, the feature computation module extracts a set of quantitative descriptors from each problem instance, capturing both fermionic and qubit-based representations, explored in detail in Section~\ref{sec:component-features}. In the fermionic representation, key quantities such as electron number, spin-orbital count, and full configuration interaction (FCI) space dimension are computed to provide an initial characterization of computational complexity. The electronic structure Hamiltonian is further analyzed through tensor factorizations, including the double-factorization (DF) decomposition, which allows for an assessment (in polynomial time) of interaction complexity and effective rank. In the qubit representation, problem instances are mapped onto Pauli decompositions via standard transformations such as Jordan-Wigner or Bravyi-Kitaev encoding, producing a Hamiltonian structured as a weighted sum of Pauli strings. From this encoding, additional complexity metrics such as Hamiltonian one-norm, Pauli string sparsity, and interaction hypergraph connectivity are computed, providing insight into the expected runtime scaling of various quantum algorithms. This feature set enables clear analysis of solver performance beyond empirical benchmarking, helping to identify trends across different classes of problem instances.

The performance analysis pipeline synthesizes results from problem instances, feature computations, and solver outputs to generate detailed benchmark reports. This module supports both direct performance comparisons and higher-level statistical analyses, incorporating standard benchmarking metrics such as solution accuracy, runtime efficiency, and resource utilization. Additionally, machine learning techniques are employed to determine solvability regions within the high-dimensional feature space, defining probabilistic boundaries between tractable and intractable problem regimes. Solvability is defined as a performer having a greater than $50\%$ chance of computing the ground state energy within chemical accuracy within each Hamiltonian's runtime requirement. These analyses, detailed in Section~\ref{sec:ML}, leverage a combination of classification models, such as support vector machines~\cite{svm} and tree-based ensembles, alongside dimensionality reduction techniques like principle component analysis (PCA) to project solver performance onto interpretable latent spaces. By integrating these methods, the benchmark does not merely provide pass/fail assessments but rather a predictive understanding of algorithmic feasibility across different problem domains.

The modular design of the benchmark accommodates a broad range of computational approaches, from traditional electronic structure methods to near-term quantum heuristics and fault-tolerant quantum algorithms. Performers can interact with the framework at multiple stages—retrieving problem instances, running solvers on selected Hamiltonians, and submitting solution files for evaluation. The repository architecture ensures that new solutions can be incorporated iteratively, enabling continuous refinement of benchmarking insights. The addition of artificial Hamiltonians with known solutions allows for controlled studies of solver behavior under varying conditions, further strengthening the benchmark’s predictive utility.

By providing a robust and extensible infrastructure for evaluating GSEE algorithms, this benchmark establishes a foundation for tracking algorithmic progress toward practical quantum advantage. The structured assessment of problem complexity, solver performance, and scalability limits offers a comprehensive view of the evolving landscape of computational quantum chemistry, guiding both theoretical advancements and experimental implementations in the field.

\subsection{Solvers Tested}
The QB GSEE Benchmark Standard Report~\cite{gsee-standard-report} currently evaluates 21 solvers, covering both classical and quantum computing approaches. Classical solvers include several variants of Selected Configuration Interaction (SHCI), with and without perturbation theory (PT) corrections, as well as Density Matrix Renormalization Group (DMRG), Coupled Cluster (CCSD and CCSD(T)), Configuration Interaction Singles and Doubles (CISD), Møller-Plesset Perturbation Theory (MP2), and Hartree-Fock (HF). These solvers were executed on high-performance classical computing hardware, such as multi-core AMD EPYC processors. 

Quantum solvers in the benchmark were based on Double-Factorized Quantum Phase Estimation (DF QPE), implemented using pyLIQTR~\cite{pyliqtr}, qualtran~\cite{qualtran}, and OpenFermion~\cite{openfermion}. These solvers estimated quantum resource requirements under various assumptions about superconducting hardware models, including different error rates and parallelization strategies. Multiple DF QPE solvers were tested, differing primarily in their truncation thresholds and orbital limits. Quantum resource estimates were calculated for each quantum solver, providing a comparison to classical methods in terms of computational feasibility.

While the full report evaluates a broad range of solvers, in this work, we focus on SHCI, DMRG, and DF QPE as representative case studies. These solvers highlight different algorithmic approaches within classical and quantum methods and serve as examples of how the repository can be used for benchmarking and solver evaluation.

\subsection{Semistochastic Heat-Bath Configuration Interaction Case Study}
\begin{figure}[htb]
    \centering
    \begin{subfigure}{0.48\textwidth}
        \centering
        \includegraphics[width=\linewidth]{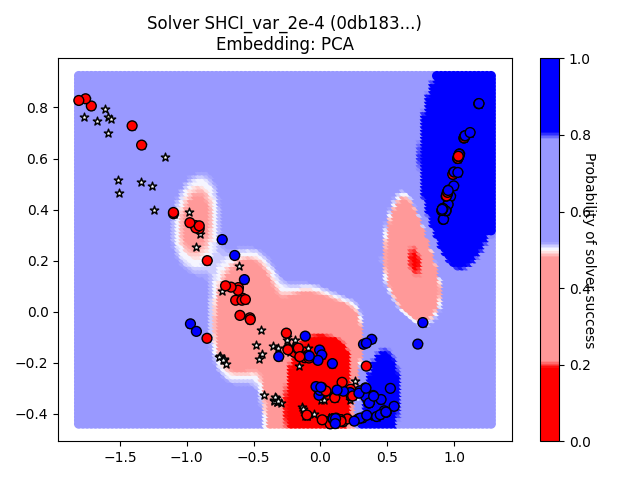}
        \caption{SHCI with $\epsilon_{\text{var}}=2\times10^{-4}$ \cite{pca_solver_3}}
        \label{fig:shci_2e-4}
    \end{subfigure}
    \begin{subfigure}{0.48\textwidth}
        \centering
        \includegraphics[width=\linewidth]{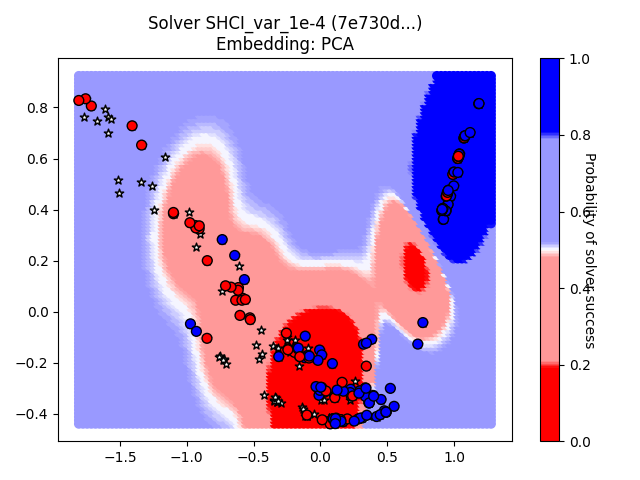}
        \caption{SHCI with $\epsilon_{\text{var}}=1\times10^{-4}$ \cite{pca_solver_4}}
        \label{fig:shci_1e-4}
    \end{subfigure}

    \vspace{0.3cm}

    \begin{subfigure}{0.48\textwidth}
        \centering
        \includegraphics[width=\linewidth]{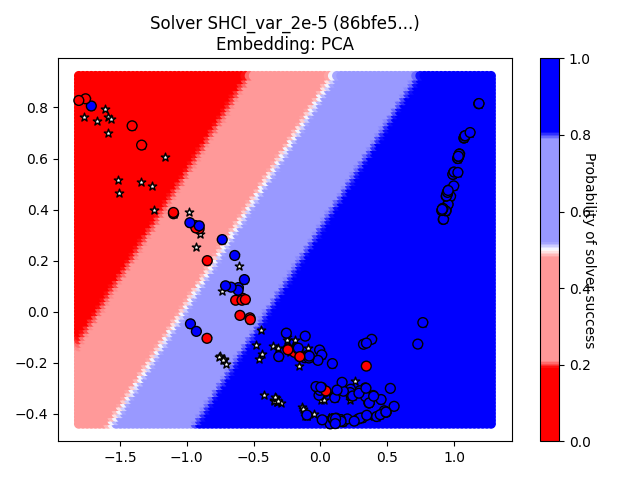}
        \caption{SHCI with $\epsilon_{\text{var}}=2\times10^{-5}$ \cite{pca_solver_2}}
        \label{fig:shci_2e-5}
    \end{subfigure}
    \begin{subfigure}{0.48\textwidth}
        \centering
        \includegraphics[width=\linewidth]{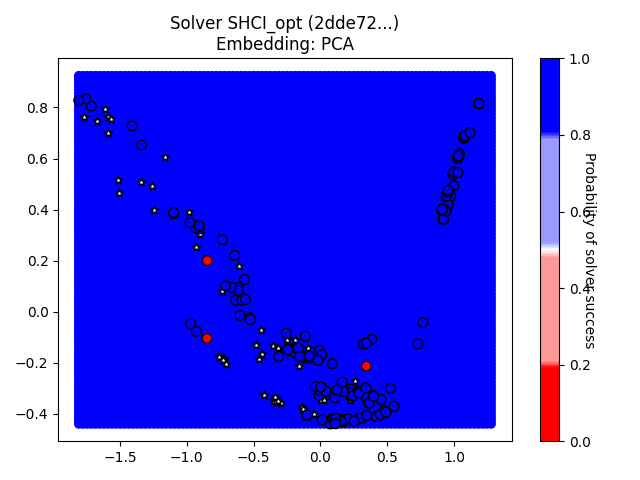}
        \caption{Optimized SHCI \cite{pca_solver_1}}
        \label{fig:shci_opt}
    \end{subfigure}

    \caption{Solvability regions for SHCI with different $\epsilon_{\text{var}}$ values and optimization settings.}
    \label{fig:shci_comparison}
\end{figure}

The solvability region demonstrates how well a specific solver performs across different problem instances. The effectiveness of the SHCI method depends on the selection threshold parameter $\epsilon_{\text{var}}$, which dictates the inclusion of determinants in the variational stage. As $\epsilon_{\text{var}}$ decreases, more terms are included, leading to higher accuracy but also increased computational costs. Figure~\ref{fig:shci_comparison} visualizes the probability of solving a given problem instance under different values of $\epsilon_{\text{var}}$, with the final plot showing the results for fully optimized SHCI, including perturbation theory (SHCI+PT) and orbital optimization.
\todo[inline]{Consider citing one or more reference on details of optimized SHCI+PT here.}

The solvability regions in Figure~\ref{fig:shci_comparison} use a color gradient to indicate the probability of success. Dark blue regions correspond to latent space where SHCI is very likely to solve problem instances, while red areas indicate areas where SHCI is likely to fail. Intermediate colors, such as light blue and pink, highlight cases where solvability is uncertain.
In addition to the color gradient, specific problem instances are marked using different symbols to distinguish their status. Blue circles represent instances that SHCI successfully solved, whereas red circles denote instances that SHCI either failed to solve or did not attempt. Guidestar problems, which are available in the dataset but lack reference energies, are marked with stars.
The axes in these figures correspond to the first two principal components obtained through Principal Component Analysis (PCA), where the horizontal axis represents the first principal component and the vertical axis represents the second. These principal components capture the key variations in problem features and solver behavior. A detailed breakdown of the feature contributions to these principal components is provided in Figure~\ref{fig:pca_components}, and further discussion on the role of PCA in solvability analysis can be found in Section~\ref{sec:ML}.

SHCI with $\epsilon_{\text{var}} = 2 \times 10^{-4}$ (Figure~\ref{fig:shci_2e-4}) displays a limited solvability region, with several red zones indicating latent space that SHCI is expected to fail to solve. Light blue and pink regions suggest uncertain success rates, where the solver struggles with marginal cases. SHCI with $\epsilon_{\text{var}} = 2 \times 10^{-5}$ (Figure~\ref{fig:shci_2e-5}) demonstrates a substantial increase in solvability, as evidenced by the expansion of the blue regions. Nevertheless, some problematic regions persist, indicating that further refinement beyond reducing $\epsilon_{\text{var}}$ is necessary to fully solve all instances. 

Interestingly, SHCI with $\epsilon_{\text{var}} = 1 \times 10^{-4}$ (Figure~\ref{fig:shci_1e-4}) solved more tasks, indicated by the filled circles changing from red to blue near $(-0.1,-0.2)$, than $2 \times 10^{-4}$, but the size of the dark blue regions decreased. Significant uncertain regions (light blue and pink zones) remain, and the transition between solvable and unsolvable areas is still broad. Optimized SHCI (SHCI Opt) (Figure~\ref{fig:shci_opt}) achieves universal solvability, as most tasks are solved
\todo[color=magenta]{Figure out why only 3 red dots but only 148 tasks solved of 226 attempted?} . Very few problem instances remain unsolved, though a small number of difficult cases persists. The improvement highlights the role of orbital optimization and SHCI+PT in refining determinant selection and enhancing energy corrections, which are essential for maximizing solvability.
\todo[inline]{This anomalous decrease in solvability in going from $\epsilon_\mathrm{var}=2\times 10^{-4}$ to $1\times 10^{-4}$, and then the worsening of the upper-left quadrant in going to $2\times 10^{-5}$: Is this reflecting a trade-off between the two solvability criteria (accuracy and runtime)? E.g., improved accuracy boosts solvability in the lower-central region, but the increased runtime kills you in the upper-left region? Is there anything worth stating on this?}

\begin{table}[th]
    \centering
    \begin{tabular}{lccc}
        \toprule
        Solver & Solvability & Tasks Solved & Tasks Attempted \\
        \midrule
        SHCI Opt (SHCI+PT with optimized orbitals) & 1.0000 & 148 & 226 \\ %2dd f1_score: [0.8333333333333334, 0.9931972789115646]
        SHCI 2e-5 & 0.6486 & 128 & 228 \\ %86b f1_score: [0.88, 0.9765625]
        SHCI 1e-4 & 0.6562 &  91 & 228 \\ %7e7 f1_score: [0.9838709677419355, 0.989010989010989]
        SHCI 2e-4 & 0.8125 &  83 & 228 \\ %0db f1_score: [0.9928057553956835, 0.9940119760479041]
        DMRG (Lowest Variational Energy)  & 0.4126 & 107 & 228 \\ %165374 f1_score: [0.9130434782608695, 0.9626168224299065]
        DF QPE    & 0.0716 &   4 & 131 \\ %2610d8 f1_score: [1.0, 1.0]
        \bottomrule
    \end{tabular}
    \caption{Solvability performance of each solver variant including how many tasks were attempted and how many were solved within both accuracy and runtime requirements.}
    \label{tab:schi_solv}
\end{table}

Table~\ref{tab:schi_solv} quantifies the solvability performance of each solver variant. SHCI Opt is the only method achieving 100\% solvability, solving enough instances in this dataset to cover the space of all instances. Lowering $\epsilon_{\text{var}}$ increases the number of tasks solved, but it only results in a 17\% improvement over SHCI 2e-5. Interestingly, this increase is sufficient to cover the full problem space.
\todo[inline]{Explicitly state how solvability is calculated for Table 1.1. It's briefly mentioned in the previous section that solvability means greater than 50\% chance of success, but at first glance of the table one might initially expect solvability to be merely tasks solved divided by tasks attempted, which it clearly isn't. Fractional solvability region for which prob of success is greater than 50\%, correct? Should also clarify that 100\% solvability doesn't mean that the solver solved all the instances. Consider a cross-ref to Section~\ref{sec:ML}.}

Decreasing $\epsilon_{\text{var}}$ increases the number of tasks solved, but it does not fully eliminate failure cases. SHCI Opt dramatically outperforms standard SHCI, highlighting the importance of orbital optimization, perturbative corrections, and extrapolation to the FCI limit. Tightening $\epsilon_{\text{var}}$ alone is insufficient — advanced solver optimizations such as SHCI Opt are essential for achieving robust performance across all instances. These findings demonstrate that while reducing $\epsilon_{\text{var}}$ helps improve SHCI’s effectiveness, further refinements through orbital optimization and perturbation corrections are necessary to achieve the highest accuracy and efficiency.
\todo[inline]{Should note that SHCI Opt is doing an extrapolation to the FCI limit, which is important (probably more than orbital opimization, but the same as PT)}

\subsection{Other Solver Comparison}
% \note{Talk about how the solvability regions change with different solvers. Hopefully show some 
% interesting insights - tbd - that demonstrate more than just pass fail (i.e., the shapes 
% of the solvability regions). Try to tie it to some commonly held notions about the 
% various solvers.}
\note{Notes from GSEE meeting: If QCs aren't able to do these things, it's just evidence that we need better algorithms/better qc are already available, such as THC or better DF.}
\begin{figure}[tb]
    \centering
    \begin{subfigure}{0.48\textwidth}
        \centering
        \includegraphics[width=\linewidth]{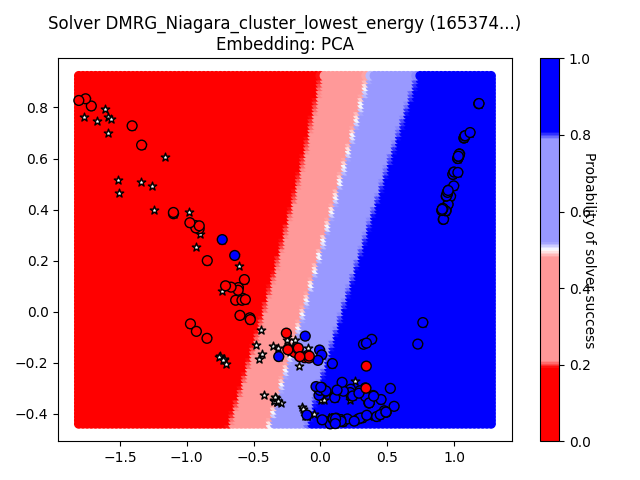}
        \caption{First Run DMRG \cite{pca_solver_6}.}
        \label{fig:dmrg}
    \end{subfigure}
    \begin{subfigure}{0.48\textwidth}
        \centering
        \includegraphics[width=\linewidth]{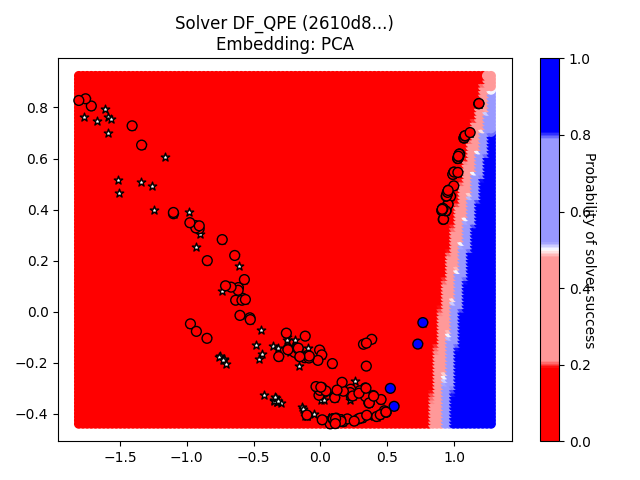}
        \caption{DF QPE \cite{pca_solver_7}.}
        \label{fig:df_qpe}
    \end{subfigure}    
    \caption{Solvability for classical and quantum methods.}
    \label{fig:pca_embeddings_extra}
\end{figure}
Comparing the solvability regions of the Density Matrix Renormalization Group (DMRG) and Double-Factorized Quantum Phase Estimation (DF QPE) provides insight into the strengths and limitations of these approaches. DMRG was run to find the lowest variational energy. \todo[inline]{JTC: We should emphasize that this is not the best performance achievable in DMRG (see discussion of the solver definition in Appendix E.2)} It exhibits a structured solvability pattern with a tighter transition region between solvable and unsolvable instances than SHCI. The method is well-suited for low-entanglement systems, which is reflected in the distinct separation of regions in Figure~\ref{fig:dmrg}. Table~\ref{tab:schi_solv} quantifies its performance, showing a solvability ratio of .4126, with 107 out of 228 tasks successfully solved.
\todo[inline]{The above statement suggests the horizontal principal-component axis is closely tied to some measure of entanglement/correlation, since the solvability regions are separated by a vertical(ish) boundary. Can more be said?}

On the other hand, DF QPE demonstrates a much smaller solvability region. Note that here `solved' means that the resource estimates are below a runtime threshold, rather than checking the DF-QPE answer against the reference energy. Many instances are marked `unsolved' because the resource estimates for many instances exceeded the runtime limitations. This suggests that quantum phase estimation in its current form struggles with many of the benchmark instances tested. As shown in Table~\ref{tab:schi_solv}, DF QPE has a significantly lower solvability ratio of .0716, solving only 4 out of 131 attempted tasks. However, this does not necessarily indicate a fundamental limitation of quantum approaches but rather highlights the need for improved algorithms and lower-cost error correction techniques. Future advancements, such as more efficient tensor hypercontraction (THC) or better double factorization strategies, could significantly enhance the performance of quantum methods.

\begin{figure}[htb]
    \centering
    \includegraphics[width=0.75\linewidth]{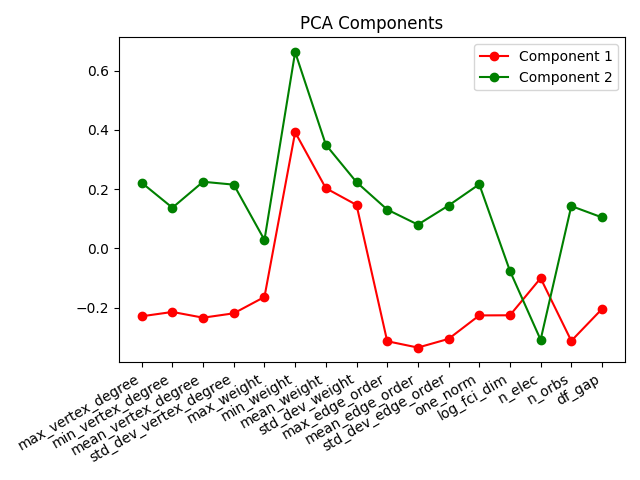}
    \caption{Principal Component Analysis (PCA) contributions for the solvability regions. The most influential features are shown for the two main components.}
    \label{fig:pca_components}
\end{figure}

\todo[inline]{Figure 5: SHAP plots / ML cluster plot / etc}
To better understand the factors contributing to the solvability regions across different solvers, we performed Principal Component Analysis (PCA) on the dataset. Figure~\ref{fig:pca_components} displays the main contributions to the PCA, highlighting which features impact the classification of solvability the most.

From the PCA results, we observe that certain graph-based and system descriptors, such as vertex degree statistics, edge order metrics, and electronic structure properties, significantly influence the classification of solvability. The first principal component is dominated by variations in molecular connectivity, while the second principal component is influenced by electronic structure parameters such as the number of orbitals and energy gaps. These insights help clarify why certain solvers perform well on specific problem instances while failing in others. 

\section{Discussion}\label{sec:discussion}
This work presents a systematic benchmarking approach for evaluating GSEE solvers across classical and quantum computing. By analyzing the solvability regions of SHCI, DMRG, and DF QPE, we have gained deeper insights into solver performance, strengths, and limitations. The results highlight that while classical methods like SHCI and DMRG demonstrate broad applicability, quantum solvers such as DF QPE require significant runtime improvements to match classical performance on benchmark tasks.

A key contribution of this work is the development of a growing database of problem instances designed to challenge and refine solver capabilities. However, it is important to note that many of the Hamiltonians in the current dataset originate from prior studies evaluating SHCI and related algorithms, creating a bias in its favor. As a result, SHCI-based methods perform particularly well on these instances, aligning with their strengths in determinant-based selection and perturbative corrections. Recognizing this limitation, future iterations of this benchmark will focus on curating a more diverse problem set, incorporating molecular systems that present significant challenges for SHCI. This expansion could include strongly correlated systems, larger multi-reference problems, and extended electronic delocalization cases, which may better highlight the comparative advantages of alternative approaches, such as tensor-network-based methods or quantum algorithms.

Beyond solver performance evaluation, this benchmarking framework has direct implications for practical use cases in scientific and industrial applications. In quantum chemistry, accurate ground-state energy estimation underpins reaction rate calculations, catalysis design, and material property prediction. The ability to systematically assess solver strengths enables researchers to select the most appropriate algorithm for a given problem. Furthermore, machine learning-assisted solvability analysis provides predictive insight into the types of problems best suited for quantum approaches, guiding the development of next-generation algorithms and hardware.

As quantum computing continues to advance, rigorous benchmarking remains essential for distinguishing theoretical speedups from practical feasibility. While our results demonstrate the current challenges facing DF QPE, they also emphasize that improvements in error mitigation, circuit optimization, and alternative ansatz constructions may significantly enhance quantum solvers' competitiveness. 

Ultimately, this work establishes a foundation for continuous benchmarking and algorithm refinement, bridging the gap between classical and quantum computational approaches. By maintaining an open, scalable benchmarking repository and incorporating emerging solvers, we can systematically track progress in GSEE and accelerate the transition toward computational methods that unlock new scientific and technological frontiers.

\note{from 2/27 GSEE meeting: this is a code that based on true and false, it can be applied to many other problems! eg optimization, cfd
could be extended to excited state information, other chemical quantities of interest
other community members could contribute. The biggest bottleneck is reference energies to get to a true/false assessment
this could use the same set of FCI dumps!
because its open source its very easy to fork and start with the tools from there}

\section{Methods}
\subsection{Ground State Energy Estimation}
%% NIB: From the hamiltonian features section (keep for consistency, but need to update language into story.
While GSEE is a fundamental problem in computational quantum chemistry and physics, systematically evaluating solver performance requires that we must precisely define the problem in a way that aligns with computational requirements and benchmarking methodologies.
Formally, The GSEE problem can be defined in the following way.
\begin{equation}
    {H = \sum_{j=0}^{K-1} \varepsilon_j \ketbra{\psi_j}{\psi_j}}
\end{equation}
where $\varepsilon_0 < \varepsilon_1 \leq \varepsilon_2 \leq \ldots \leq \varepsilon_{K-1}$ are the eigenvalues of $H$ and $\{\ket{\psi_j}\}$ are the set of orthonormal eigenstates of $H$.
The goal is to estimate the lowest eigenvalue $\varepsilon_0$, corresponding to the system's ground-state energy.

In practical applications, we often consider a promise problem formulation where there exists an `easy-to-prepare' state $\rho$ (of the same dimension as $H$) with a nonzero overlap with the true ground state $\ket{\psi_0}$, and a guaranteed spectral gap $\varepsilon_1 - \varepsilon_0$. Namely, we assume there exist two parameters, $\xi \in (0,1)$ and $\Delta > 0$, such that:
\begin{equation}
    \bra{\psi_0} \rho \ket{\psi_0} \geq \xi, \quad \varepsilon_1 - \varepsilon_0 \geq \Delta.
\end{equation}
The task is to estimate $\varepsilon_0$ with accuracy $\gamma$ and confidence $1 - \delta$, meaning that the output energy estimate ${E}_0$ must satisfy the failure probability:
\begin{equation}
    P(E_0 - \varepsilon_0 > \gamma) < \delta,
\end{equation}
for given small $\gamma > 0$ and $\delta \in (0,1)$.
% We are often interested in a promise problem formulation, such that there exists some `easy-to-prepare' state $\rho$ (of the same dimension as $H$) with at least some overlap with the true ground state $\ket{\psi_0}$, and that the spectral gap $\varepsilon_1 - \varepsilon_0$ is at least some nonzero value.
% Let $p_j := \bra{\psi_j} \rho \ket{\psi_j}$ be the overlap between $\rho$ and $\ket{\psi_j}$ for $j \in\mathbb{Z}_{K}$.
% We assume that two numbers $\xi \in (0, 1)$ and $\Delta > 0$ are given such that $\bra{\psi_0} \rho\ket{\psi_0} \geq \xi$ and $\varepsilon_1 - \varepsilon_0 \geq \Delta$.
% The problem is to estimate $\varepsilon_0$ with accuracy $\gamma$ and confidence $1 - \delta$, i.e., to output a sample from a random variable ${E}_0$ such that the failure probability satisfies $P(E_0 - \varepsilon_0 > \gamma) < \delta$ for given small $\gamma > 0$ and $\delta \in (0, 1)$.

In computational quantum chemistry, the electronic structure Hamiltonian is can be represented using an FCIDUMP file. The FCIDUMP is an ASCII file that encodes the one- and two-electron integrals in a molecular orbital basis, providing all the necessary information to construct a full configuration interaction (FCI) Hamiltonian. Unlike experimental references or molecular specifications, FCIDUMP files specify the computational problem directly, making them an essential resource for algorithm benchmarking. By providing a structured, reproducible representation of problem instances, FCIDUMP files enable consistent evaluation of solver performance across classical and quantum methods.

To assess solver efficiency and scalability, it is crucial to analyze problem features derived from these Hamiltonians. Section~\ref{sec:component-features} introduces key Hamiltonian descriptors, such as electron count, basis set size, and interaction complexity, which serve as the foundation for our benchmarking framework. These features allow for systematic comparisons of solver performance across a diverse set of problem instances, ultimately guiding the development of more efficient ground-state energy estimation techniques.

\subsection{Hamiltonian Features}\label{sec:component-features}
To systematically benchmark GSEE solvers, we extract key Hamiltonian features that characterize problem complexity and solver performance. These features are derived from two complementary representations of the Hamiltonian: (i) the \textit{fermionic representation}, which directly encodes the electronic structure problem, and (ii) the \textit{qubit representation}, which describes the problem in terms of quantum computational resources. 

In designing these features, we ensured that they could be computed efficiently, enabling their evaluation on large-scale problem instances. By implementing feature computation methods that run in polynomial time, we facilitate benchmarking on significantly larger Hamiltonians than would otherwise be feasible. This scalability allows for a more comprehensive and representative assessment of solver performance across a broad range of problem complexities.

\subsubsection{Fermionic Representation Features}
The fermionic representation expresses the electronic structure Hamiltonian in second quantization:
\begin{equation}
    H = \sum_{i,j} h_{ij} a_i^\dagger a_j + \frac{1}{2} \sum_{i,j,k,l} h_{ijkl} a_i^\dagger a_k^\dagger a_l a_j.
\end{equation}
where $a_i^\dagger$ and $a_j$ are fermionic creation and annihilation operators, and $h_{ij}, h_{ijkl}$ encode one- and two-body interactions.

Several key features are derived from this representation, providing insight into the structure and computational complexity of the problem. The number of electrons, denoted as $\eta$, and the number of spin-orbitals, $N$, define the size of the problem and directly influence the dimensionality of the underlying Hilbert space. A more refined metric of complexity is given by the logarithm of the Full Configuration Interaction (FCI) space dimension, which captures the number of possible electron configurations in the chosen basis and serves as an upper bound on the classical computational cost of solving the system exactly.

Another important set of features is obtained through the double-factorization (DF) decomposition, which approximates the two-electron integral tensor with a reduced-rank representation. The DF rank, denoted as $L$, quantifies the number of significant terms in this decomposition, thereby influencing computational scaling in both classical and quantum solvers. Additionally, the DF eigenvalues describe the spectral distribution of these terms, while the DF eigenvalue gap, defined as the difference between the two largest eigenvalues, provides a measure of whether a Hamiltonian is dominated by a single term or requires multiple components for an accurate description.

Further mathematical details, including derivations and the role of DF decomposition, are provided in Appendix~\ref{app:ham_features}.

\subsubsection{Qubit Representation Features}
Quantum algorithms solve GSEE by mapping fermionic Hamiltonians to qubit-based representations using transformations such as Jordan-Wigner or Bravyi-Kitaev encodings. In this framework, the Hamiltonian is expressed as a weighted sum of Pauli strings:
\begin{equation}
    H = \sum_{e \in E} h_e P_e, \quad P_e \in \{I, X, Y, Z\}^{\otimes n}.
\end{equation}

Several features characterize the complexity of this representation. The Hamiltonian one-norm, denoted as $\lambda(H)$, is an important quantity that determines the scaling of quantum algorithms based on phase estimation and linear combination of unitaries, as it sets a fundamental bound on the simulation time required to evolve the system. Another critical property is the number of Pauli strings, $|E|$, which reflects the sparsity of the Hamiltonian and directly impacts quantum circuit depth and gate complexity.

Additionally, the structure of the Hamiltonian can be analyzed in terms of an interaction graph, where each Pauli term is treated as an edge connecting the qubits it acts upon. The statistical properties of this graph, including vertex degrees, edge orders, and edge weight distributions, provide further insights into entanglement structure, qubit connectivity, and the expected difficulty of simulating the system using quantum algorithms.

These features collectively determine the scalability and feasibility of different computational approaches. A full discussion of their computational relevance is provided in Appendix~\ref{app:ham_features}.

\subsubsection{Feature Summary and Correlation}
\begin{figure}[t]
    \centering
    \includegraphics[width=0.755\linewidth]{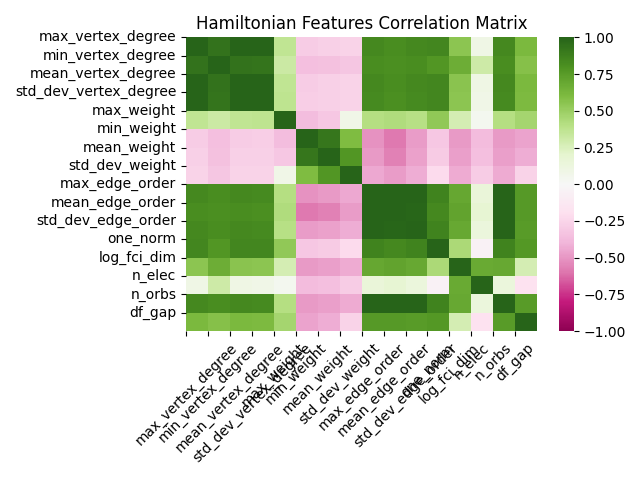}
    \caption{Hamiltonian features correlation matrix~\cite{hamiltonian_correlation}. The color intensity represents the degree of correlation between features, with highly correlated features appearing in darker green shades while anti-correlation is shown in shades of pink.}
    \label{fig:featcor}
\end{figure}

To better understand the relationships between different Hamiltonian features, Figure~\ref{fig:featcor} shows their correlation matrix across current problem instances. This helps identify which features provide unique information and which are redundant.

Figure~\ref{fig:featcor} reveals that some features are strongly correlated. For example, maximum and minimum vertex degree show a high correlation, meaning they likely capture similar information about solver performance. On the other hand, some features, like the number of electrons ($n_{\text{elec}}$), have little correlation with qubit-based features such as edge order statistics or Pauli string count.

Since many features overlap in the information they provide, future work could reduce the feature set without losing predictive power. Removing redundant features would also create space for additional complexity measures to further improve the benchmarking framework.

Table~\ref{table:feature_table} summarizes the primary Hamiltonian features, categorized by representation type.

\begin{table}[ht]
\centering
\begin{tabular}{lll}
    \toprule
    \multicolumn{3}{c}{Table of Selected GSEE Problem Features} \\
    \midrule
    Representation     & Feature                        & Notation / Formula \\
    \midrule
    Fermionic         & Number of electrons            & $\eta$ \\    
                      & Number of spin-orbitals        & $N$ \\    
                      & Log FCI size                   & $\log_{10} \left( \binom{N}{N_{\alpha}} \binom{N}{N_{\beta}} \right)$ \\    
                      & DF eigenvalues                 & $\{\lambda_\ell\}_{\ell =0}^{L-1}$ \\    
                      & DF rank                        & $L$ \\    
                      & DF eigenvalue gap              & $|\lambda_0 - \lambda_1|$ \\    
    \midrule
    % Qubit            & Number of qubits               & $n$ \\    
    Qubit            & $H$ one-norm                    & $\lambda(H)$ \\    
                     & Number of Pauli strings         & $|E|$ \\    
                     & Edge order stats                & $\mathrm{ord}(e)$ \\    
                     & Vertex degree stats             & $\mathrm{deg}(v)$ \\    
                     & Edge weight stats               & $h_e$ \\    
    \bottomrule
\end{tabular}

    \caption{Summary of key GSEE problem features. \textit{Fermionic features}: $\eta$ and $N$ are problem size parameters of the GSEE problem. Together, they define the log FCI size of the ground state wavefunction, which captures the number of configurations of $\eta$ electrons in $N$ spin-orbitals where $N_\alpha$ and $N_\beta$ are the number of spin-up and spin-down electrons respectively, and $N_\alpha + N_\beta = \eta$. The double factorization decomposition of the chemical Hamiltonian leads to a set of $L$ eigenvalues $\{\lambda_\ell\}_{\ell =0}^{L-1}$ that provide a metric of the Hamiltonian's complexity. Specifically, we look at the total number of them $L$ and the gap between the 1st and 2nd largest eigenvalues $|\lambda_0 - \lambda_1|$. \textit{Qubit features}: In the Qubit representation, a chemical Hamiltonian is mapped to a Pauli Hamiltonian $H = \sum_{e \in E} h_e P_e$ with an associated interaction hypergraph $G=(V,E)$ of $n$ vertices (system qubits) with hyperedges that track the Pauli strings $P_e \in \{I,X,Y,Z\}^{\otimes n}$. From this representation, the induced Hamiltonian's one-norm $\lambda(H)=\sum_{e \in E} |h_e|$, the number of Pauli strings $|E|$, and graph statistics include the max, min, mean, and standard deviations of graph quantities such as edge orders, vertex degrees, and edge weights can be derived.}
    \label{table:feature_table}
\end{table}

\subsection{Problem Selection}\label{sec:problem-selection}
The database comprises a curated collection of problem instances spanning diverse chemical and physical challenges, structured to facilitate benchmarking and algorithm development. These instances are categorized into benchmark instances, which have known reference solutions derived from classical computational methods, and guidestar instances, which are complex problems exceeding current computational capabilities but hold practical value if solved.

A key requirement for benchmark instances is the availability of reference energies, essential for validating solutions and assessing algorithm performance. These energies are sourced from literature, self-computed, or derived from artificially constructed Hamiltonians with planted solutions~\cite{wang_planted_2025}.  While we aimed for all reference energies to be within chemical accuracy, this was not fully achievable in every case. Instead, we prioritized a broad and diverse set of instances over perfect reference values, ensuring a dataset that supports a wide range of solver evaluations. If variational methods produce energies lower than the extrapolated reference values, sharing these results with the community would be valuable for further benchmarking.

The chemistry included in the benchmark spans a diverse set of molecular systems that represent significant computational challenges. Small diatomic molecules such as Ne\(_2\) and LiH serve as well-characterized reference systems, while larger organic molecules, including propane (C\(_3\)H\(_8\)) and benzene, present more intricate electronic structures that test computational methods~\cite{gao2024distributed, eriksen2020benzene}. Transition metal compounds, exemplified by the chromium dimer (Cr\(_2\)), are included due to their long-standing reputation as grand challenges in computational chemistry, driven by strong electron correlation effects~\cite{larsson2022cr2}. Atmospheric chemistry is represented by ozone (O\(_3\)), an environmentally significant molecule whose multiple geometric configurations pose distinct computational challenges~\cite{chien2018excited}. 

Beyond molecular chemistry, the database includes problem instances relevant to condensed matter and strongly correlated electron systems. The inclusion of model Hamiltonians, such as the Fermi-Hubbard and Heisenberg models, enables controlled studies of quantum many-body interactions and benchmarking of solvers on systems with known theoretical properties~\cite{Wu2024_vscore}. Additionally, utility-driven problems such as homogeneous catalysis, corrosion modeling, and protein interactions provide relevance to industrial and biochemical applications~\cite{bellonzi2024, Nguyen2024, otten2024quantum}. These problem instances align with real-world computational challenges where quantum advantage may eventually emerge. Further, planted solutions based on homogeneous catalysts~\cite{wang_planted_2025} provide industrial scale systems with exactly known reference energies.

\begin{figure}[ht]
    \centering
    \includegraphics[width=0.75\linewidth]{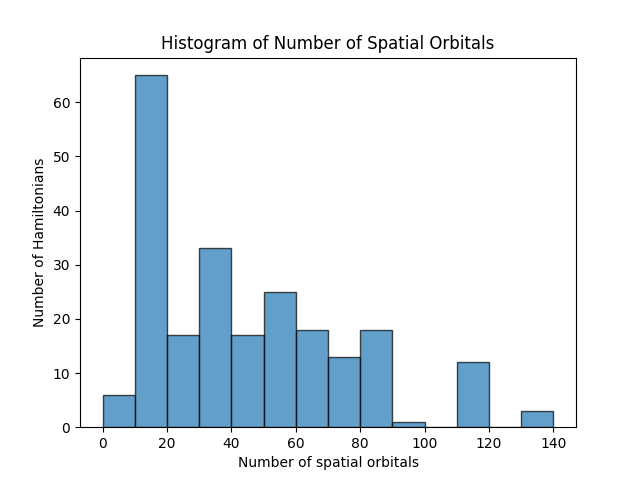}
    \caption{Histogram of the number of spatial orbitals in the problem instances, showing the distribution of Hamiltonians by orbital count~\cite{num_orbitals_histogram}.}
    \label{fig:num_orbitals}
\end{figure}

To ensure consistency and usability, each problem instance adheres to a standardized schema, incorporating metadata, problem classification, supporting information, and computational requirements. As shown in Figure~\ref{fig:num_orbitals}, the problem set spans a range of orbital sizes, allowing for evaluations across varying levels of computational complexity. This structure enables systematic evaluation and comparison of algorithms while supporting reproducibility.

New instances can be added through curated contributions or artificially generated Hamiltonians, ensuring continuous expansion of the benchmark dataset. The problem instances serve as a foundational resource for testing and refining computational methods, particularly in quantum chemistry and condensed matter physics, with the potential for broader scientific and engineering applications.

\subsection{Predicting Solvability Using Machine Learning}\label{sec:ML}
Given the problem features for a Hamiltonian and a performer’s solution file detailing their algorithm’s ability to solve it, we use machine learning (ML) techniques to predict solvability for novel test points. This allows us to estimate the \textit{solvable space}, the region in feature space where a given solver is expected to succeed. Using a trained ML model, we define a decision boundary between classically solvable and intractable regions, enabling us to estimate the proportion of the problem space that the performer’s method can handle.

Since the problem feature space is high-dimensional and sparse, we transform it into a lower-dimensional \textit{latent space} before performing solvability computations. Initially, we implemented Non-Negative Matrix Factorization (NNMF) to construct an interpretable, invertible latent representation that preserves the non-negativity of physical Hamiltonian features. However, we ultimately chose Principal Component Analysis (PCA) for its greater interpretability and numerical stability while still retaining NNMF as an alternative option in the repository. PCA efficiently captures variance in the data while maintaining a straightforward mapping between original and latent feature spaces~\cite{PCA_ref}.

A Support Vector Machine (SVM) classifier is trained on benchmark data to distinguish between solvable and unsolvable problem instances. Once trained, the model is used to generate 10,000 novel test points in the latent space and predict their probability of being solvable. The solvability ratio is then computed as the fraction of points exceeding a predefined probability threshold of 0.5. This method also includes classification metrics, probability mapping, and feature importance analysis. Further details on this methodology, including latent space computation and model selection, are provided in Appendix~\ref{app:SSE}.

\section*{Acknowledgments}
This work was funded under the DARPA Quantum Benchmarking program. 
The authors would like to thank K. Morrell for invaluable assistance with using pyLIQTR's implementations of double-factorized phase estimation.
This research was partly enabled by Compute Ontario (computeontario.ca) and the Digital Research Alliance of Canada (alliancecan.ca) support. Part of the computations were performed on the Niagara supercomputer at the SciNet HPC Consortium. SciNet is funded by Innovation, Science, and Economic Development Canada, the Digital Research Alliance of Canada, the Ontario Research Fund: Research Excellence, and the University of Toronto. 
J.T.C. would like to thank A.M. Rey for her hospitality during his visit to JILA at CU Boulder. %Note: JILA is not an acronym.

\pagebreak
\appendix
% \input{909_all_pcas_plot}

% \section{Introduction}
% \input{010_intro}

% \section{Framework Overview}
% \input{011_benchmark_components}

% \section{User Workflows and Instructions}
% \input{012_users}

% \section{Problem Instance Database}
% \input{021_problem_instances}

\section{Hamiltonian Features: Extended Details} \label{app:ham_features}

% \subsection{Formal Treatment}

% 

In standard classical and quantum complexity theory, it is typical to consider the scaling of a problem relative to a single input parameter, usually referred to as its problem size. The traditional approach in computer science is to look at the scaling of a solver (algorithm) with problem size to assess the tractability of the associated class of problems. However, it is usually the case that the problem description contains a range of parameters that specify the problem that may have complicated dependence on the traditional notion of size; this is especially the case with quantum algorithms for solving GSEE problems \cite{Bremner2022}. This motivates the procurement of a set of \textit{problem features} that capture GSEE problem hardness in some practical and meaningful way. The goal is to define the characteristic quantities that, when combined, differentiate easy and difficult instances to the greatest extent possible.

Problem features are quantities derived from the mathematical description of a GSEE problem instance.
%% NIB: moved to GSEE def
% The GSEE problem can be formally defined in the following way.
% \begin{equation}
%     {H = \sum_{j=0}^{K-1} \varepsilon_j \ketbra{\psi_j}{\psi_j}}
% \end{equation}
% where $\varepsilon_0 < \varepsilon_1 \leq \varepsilon_2 \leq \ldots \leq \varepsilon_{K-1}$ are the eigenvalues of $H$ and $\{\ket{\psi_j}\}$ are the set of orthonormal eigenstates of $H$.
% We are often interested in a promise problem formulation, such that there exists some `easy-to-prepare' state $\rho$ (of the same dimension as $H$) with at least some overlap with the true ground state $|\psi_0\rangle$, and that the spectral gap $\varepsilon_1 - \varepsilon_0$ is at least some nonzero value.
% % Let $p_j := \bra{\psi_j} \rho \ket{\psi_j}$ be the overlap between $\rho$ and $\ket{\psi_j}$ for $j \in\mathbb{Z}_{K}$.
% We assume that two numbers $\xi \in (0, 1)$ and $\Delta > 0$ are given such that $\bra{\psi_0} \rho\ket{\psi_0} \geq \xi$ and $\varepsilon_1 - \varepsilon_0 \geq \Delta$.
% The problem is to estimate $\varepsilon_0$ with accuracy $\gamma$ and confidence $1 - \delta$, i.e., to output a sample from a random variable ${E}_0$ such that the failure probability satisfies $P(|E_0 - \varepsilon_0| > \gamma) < \delta$ for given small $\gamma > 0$ and $\delta \in (0, 1)$.
Problem features of GSEE instances, or of any other computational problem, can only be completely rigorously defined when they are analyzed with a specific algorithm and representation in mind. In this case, the relevant problem features are precisely the scaling parameters of the algorithm, appropriately made independent of one another. Such a rigorous definition also includes full specification of any compression schemes, symmetry reductions, and any other procedure. For example, there are many classes of Hamiltonians that are efficient to solve i.e. produce the ground state energy exactly through some analytic or classically computable means. When analyzed with respect to a completely general GSEE solver / algorithm e.g. full quantum phase estimation, the appropriate problem features for these easy-to-solve instances are the same as for any other instance - in the sense that the quantum circuit generated by the general procedure will scale with the underlying parameters of the algorithm. However, applying e.g. a compression that is applicable to this subclass may result in dramatically shorter circuits and thus including such a preprocessing step delineates a separate algorithm. Our goal is to proceed by first examining specific quantum algorithms and their parameterized scaling for GSEE, and slowly remove constraints on pure rigor as we make some trade off toward practical and meaningful problem features.

\subsection{Fermionic Representation Features}

In the context of electronic structure theory, the Hamiltonian is derived from the geometry of a chemical system. The electronic structure Hamiltonian in an arbitrary second-quantized basis can be expressed as
\begin{equation}\label{eq: second-quantized ham GTO}
    H = \sum_{i,j} h_{ij} \  a^\dagger_i a_j + \frac{1}{2}\sum_{i,j,k,l} h_{ijkl} \  a^\dagger_i a^\dagger_k a_l a_j,
\end{equation}
where $a^\dagger_i$ and $a_j$ are fermionic creation and annihilation operators associated with spin-orbitals $i,j \in \mathbb{Z}_N$. The coefficients $h_{ij}$ and $h_{ijkl}$ are the one- and two-body integrals derived from electron-nuclei and electron-electron interactions, respectively; these are stored in the problem instance database as FCIDUMP files. A subset of the features we compute is derived from this fermionic representation of the Hamiltonian. These features are listed in Table \ref{table:feature_table} under the representation category of \textit{Fermionic}. The \textit{number of electrons}, $\eta$, and the \textit{number of spin-orbitals}, $N$, are problem size features that determine the size of the GSEE problem (the dimensionality of the encoded electronic Hamiltonian). In reality, the Hilbert space of even hydrogen-like atoms with one electron has countably infinite dimension, and for computational purposes we must restrict ourselves to some electron state subspace spanned by some choice of $N$ spin-orbitals. The degree to which a chosen basis set approximates the low energy subspace of the full Hamiltonian is key and motivates modern research into useful basis sets for different systems, like the dual plane wave basis \cite{Babbush2017LowDepthQS}, gausslets \cite{White2017HybridGS}, etc. However, for a given truncation into a finite subspace spanned by the $N$ spin-orbitals, more electrons does not necessarily translate to problem hardness. 
For a molecule with $\eta$ electrons and a given truncation to $N$ spin-orbitals, there are $(\substack{N \\ \eta})$ relevant Slater determinants (or configurations) that form a basis of the $\eta$-electron Hilbert space. This is maximized at half-filling ($\eta \approx N/2$) and motivates the feature \emph{log FCI size}, which is just the logarithm of this fixed-particle-number dimension \cite{Szab1982ModernQC}. When taken base 2, the log FCI size has the advantage of giving the information theoretic minimum number of qubits that generate a space of the same size; base 10 is also standard.

In order to gain further insights into the structure and complexity of fermionic Hamiltonians, Eq.~\eqref{eq: second-quantized ham GTO} can be decomposed in a variety of ways. One method we examine in particular is the so-called double-factorization (DF) procedure \cite{df_trotter}. This procedure involves a tensor factorization of $h_{ijkl}$ into a set of scalars $\lambda_\ell$ and matrices $g^{(\ell)} \in \mathbb{R}^{N \times N}$ for $\ell \in \mathbb{Z}_L$. Eq.~\eqref{eq: second-quantized ham GTO} can then be rewritten as 
\begin{equation}
    \label{eq:df-ham}
H = \sum_{ij} h_{ij} \ a^\dagger_i a_j + \frac{1}{2} \sum_{\ell = 0}^{L-1} \, \lambda_\ell \, \Big(\sum_{ij} [g^{(\ell)}]_{ij} a^\dagger_i a_j \Big)^2.\\
\end{equation}
Such that the $N^4$-many quartic fermionic terms in Eq. (\ref{eq: second-quantized ham GTO}) are grouped and decomposed into a linear combination of $L$-many ``fragments'' in equation (\ref{eq:df-ham}), each of which is a squared quadratic Hamiltonian determined by the corresponding coefficient matrix $g^{(\ell)}$.
From this tensor factorization, we derive the following features: 1) \textit{DF rank}, 2) \textit{DF eigenvalues}, and 3) \textit{DF eigenvalue gap}. 

Each of the $L$-many fragments determined by $g^{(\ell)}$ are individually efficient to solve via orbital rotation. However, the linear combination of $L$ of these squared terms makes the total Hamiltonian difficult (empirically known to be $L = O(N)$ for real chemical systems \cite{df_trotter}, but is bounded above by $O(N^2)$). We can motivate these features most easily in a Trotterization based algorithm. It can be shown that the complexity of a Trotter step scales linearly with the DF rank, and that the distribution of eigenvalues give a picture on the number of large (non-perturbative) fragments that contribute to the overall Hamiltonian. Take the DF eigenvalues in descending order, such that $\lambda_0 \geq \lambda_1 \geq \ldots \geq \lambda_{L-1} $. Then we can interpret the DF eigenvalue gap $| \lambda_0 - \lambda_1|$ as a measure of how much the second largest fragment contributes relative to the largest fragment. If this gap is very large, that means that the second largest fragment is essentially just a perturbation on the largest fragment (and therefore so are all the others). If the gap is very small, then it is a very poor approximation to say that the Hamiltonian is dominated by a single fragment, which is intuitively harder to solve. 

\subsection{Qubit Representation Features}

\begin{figure}[ht]
    \centering
    \includegraphics[width=.55\linewidth]{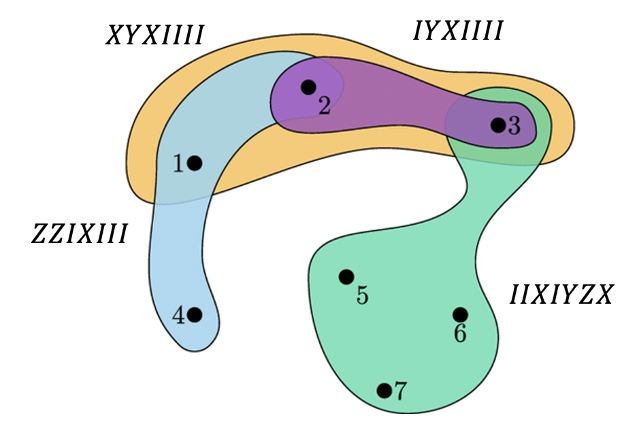}
    \caption{Cartoon of the interaction hypergraph $G=(V,E)$ associated with a Pauli Hamiltonian acting on $n=7$ qubits $H = h_{e_1} ZZIXIII + h_{e_2} XYXIIII + h_{e_3} IYXIIII + h_{e_4} IIXIYZX$ for some edge weights $h_{e_i}$ for $e_i \in E$. Figure adapted from \cite{Schawe2022}.}
    \label{fig:pauli_string_hypergraph}
\end{figure}

There are various quantum approaches to solving the GSEE problem using quantum computers, ranging from different flavors of QPE, such as qubitized QPE \cite{Babbush2018_qubitizedpe, thc_complexity_qpe, Nguyen2024, ozone_df_reference, bellonzi2024}, to statistical QPE techniques \cite{linlin_earlyfaulttolerant, Lin2021HeisenbergLimitedGE, Wang2023_earlyfaulttolerant}, the recently developed method QPE without controlled unitaries \cite{clinton2024_uncontrolled_qpe}, and various other methods like quantum Krylov subspace techniques \cite{lee2024_krylov, Cortes2021QuantumKS}, including quantum filter diagonalization \cite{Parrish2019QuantumFD}. However, quantum computers are not naturally designed to work directly with fermionic Hamiltonians. Instead, they operate using a register of $n$ qubits, whose state vectors span an equivalent Hilbert space $ \mathcal{H}_n\coloneqq\text{Span}\{\ket{0},\ket{1}\}^{\otimes n}$. Within this framework, the $4^n$ Pauli strings form an orthogonal basis in the real vector space $L_H(\mathbb{C}^{2^n})$, which is the space of Hermitian operators acting on the $n$-qubit Hilbert space. In this representation, the Hamiltonian can be expressed as $H = \sum_{P} h_P P$ where $h_P$ is the coefficient of the Pauli string $P \in \{I,X,Y,Z\}^{\otimes n}$. The structure of these Pauli strings can be visualized as an interaction hypergraph $G = (V, E)$, where vertices correspond to qubits and hyperedges represent Pauli strings acting on subsets of qubits. A cartoon of this interaction graph is shown in Figure \ref{fig:pauli_string_hypergraph}. The transformation from a fermionic Hamiltonian to a Pauli string basis can be done in several ways. The Jordan-Wigner transformation \cite{Jordan1928berDP} gives a set of mutually anti-commuting Pauli strings that can be used to faithfully represent the canonical anti-commutation relations of the fermionic ladder or Majorana operators. Although this is the most widely used due to its simplicity, the choice of embedding fermionic operators in the Pauli string representation is highly degenerate, and is itself a subroutine that must be specified for a complete description of an algorithm for the fermionic GSEE problem. Other popular encodings include the Bravyi-Kitaev transformation \cite{Bravyi2000FermionicQC}, ternary trees encoding \cite{Jiang2019OptimalFM}, as well as symmetry-aware encodings, which represent resource optimization procedures such as the locality-preserving BKSF encoding \cite{Setia2018SuperfastEF}, and possibly others like the Sierpinski encoding \cite{Harrison2024AST} or those specific to symmetries within certain lattice models \cite{Derby2021_compactmappings_1, Derby2021_compactmappings_2}. These encodings have historically been developed to minimize the average \emph{Pauli weight} of the terms in the encoded Hamiltonian, first being $O(N)$ in Jordan-Wigner, to $O(\log(N))$ in the case of Bravyi-Kitaev, and eventually independent of system size for the highly structured lattice cases. We emphasize that any features we discuss that look explicitly at the Pauli string representation are entirely dependent on the choice of encoding at this step. Indeed, there are encodings in second quantization which make use of particle conservation of the electronic structure Hamiltonian, which use fewer than $N$ qubits \cite{kirby_secondquant, harrison2023reducingqubitrequirementjordanwigner, Shee_qubit_efficient, Moll_2016, carolan2024succinctfermiondatastructures}.
The Pauli string features we define are only comparable with molecular GSEE instances that \emph{use the same encoding}, and this highlights again the notion that no problem feature is guaranteed to be invariant under different choice of efficient representation or algorithm subroutines. For this current work, we define molecular Pauli string instances using the Jordan-Wigner transformation.  

Once a Pauli string representation for a given fermionic Hamiltonian has been chosen (e.g., a specific encoding has been used), all quantum algorithms for solving the GSEE problem depend entirely on the collective structure of the Pauli string terms. For instance, quantum simulation techniques such as linear combination of unitaries (LCU), qDRIFT, and qubitization scale with respect to the 1-norm of the decomposed Hamiltonian. The 1-norm is closely tied to the specific decomposition technique used, which in turn depends on various optimization and reduction strategies \cite{Patel2024_bliss, Koridon2021_orbrot, Loaiza2023, yen2022_grouping_paulis,boyd2024_grouping_paulis}. These strategies utilize features of the Pauli strings, such as grouping anti-commuting Pauli strings together to enhance efficiency. In contrast, quantum simulation through Trotter-based techniques is independent of the 1-norm; however, their complexity is influenced by the Trotter errors, which are tied to the decomposed Hamiltonian, particularly the commutativity of the Pauli strings \cite{Childs2021_theoryoftrotter}. Hence, it is possible to evaluate various structural properties of this Hamiltonian by examining a hypergraph that is derived from the interaction Pauli. Using that approach, we derive the following Pauli string features: 1) \textit{Number of qubits}, 2) \textit{H one-norm},  3) \textit{Number of Pauli strings}, 4) \textit{Edge order stats}, 5) \textit{Vertex degree stats}, 6) \textit{Edge weight stats}. These features are
listed in Table \ref{table:feature_table} under the representation category of \textit{Qubit}.

% \todo[inline]{The Pauli Hamiltonian interaction hypergraph figure isn't referenced anywhere in the text.} %% SOLVED

% \todo[inline]{Is number of qubtis always 2 * number of orbitals? It seems like it should be, which makes it perhaps a bad feature to use in the machine learning set, but I don't think we're actually using it? It isn't in the PCA plot?, figure 1.3  -> I am guessing the original thought about the number of qubits is that it might refer to the number of qubits used in the algorithm, which can be more than twice the number of qubits, and this varies depending on the implementation. But I am going to remove the number of qubits from the feature.}
%The \textit{number of qubits}, $n$, plays a fundamental role in determining the memory requirements or space complexity of the quantum simulation. This is true regardless of whether the simulation is to be performed on either a quantum or classical computer \cite{Loaiza2023, Childs2012, Chapman2023_freefermions}. 
The \textit{Hamiltonian one-norm} determines the scale of the Hamiltonian in terms of its coefficients. In the context of quantum simulations, when the Hamiltonian is encoded using a linear combination of unitaries (LCU), the time complexity of the encoding scales with the one-norm of the Hamiltonian \cite{Loaiza2023, Childs2012}. 
The \textit{number of Pauli strings}, $|E|$, reflects the sparsity of the Hamiltonian and directly impacts the time complexity of algorithms like LCU. Additionally, the number of Pauli strings dictates the number of ancillas needed for implementing the LCU protocol, making this a key factor in understanding the overall resource requirements for quantum simulations \cite{Loaiza2023, deVeras2022}.  
The \textit{edge order} or Pauli weight of a given term, denoted as $\mathrm{ord}(e) = k(e)$, quantifies the complexity of the interactions between the terms in the Hamiltonian. It tracks the number of non-identity terms present over all Pauli strings in $H$ and is related to the sparsity of the Hamiltonian. Additionally, the time complexity of implementing LCU scales with the mean edge order as the complexity of evaluating each term is directly proportional to the number of non-identity terms it has. 
Similarly, the \textit{vertex degree}, $\mathrm{deg}(v) = |{ e \in E : v \in e }|$, measures how connected the qubits are within the Hamiltonian. This provides insight into the entanglement structure of the system. Hence, vertex degree statistics are another indicator of Hamiltonian sparsity and, consequently, the difficulty of encoding and simulating the Hamiltonian using quantum algorithms. Finally, the \textit{Edge weight stats} or distribution of edge weights, $h_e$, can significantly influence the difficulty of the Hamiltonian simulation. Evidence suggests that Hamiltonians with a low variance in their edge weights are more easily solvable. In particular, if the absolute values of all edge weights are equal, say $|h_e| = 1$ for all edges $e \in E$, the Hamiltonian may exhibit properties that make it easier to solve, as seen in certain classes of sparse Hamiltonians \cite{Chen2024_quantumlyeasy}. This feature highlights how even small changes in the distribution of edge weights can drastically affect the solvability of the Hamiltonian.

Once the features of the GSEE problems of interest are computed, these define $D$-dimensional vectors in \textit{feature space} $\mathbf{x}(H) \in \mathbb{R}^D$ where each entry of the vector is the value of a specific feature associated with the Hamiltonian $H$, where $D$ represents the total number of selected features. Our goal is to determine, for a given choice of algorithm implemented on a specific hardware platform, the regions of solvability and their boundaries. That is, we want to define a collection of regions in feature space that capture subsets of metric values for which the solver in consideration is \textit{likely} to successfully solve the problem. The resolution and accuracy of determining these solvability regions is tied to our ability to construct benchmark problems for which we know the solution a priori, for example \cite{Kojima2024_orbitalrotatedfermihubbardmodelbenchmarking}. Successfully solving benchmark problems informs a machine learning model that estimates the solvability regions which provides a measure of progress towards a hardware platform that is capable of solving GSEE problems that carry significant scientific and industrial utility e.g., those that are currently outside of what is possible with classical hardware \cite{Menczer2024}.

\section{Machine Learning: Solvable Space Estimation}\label{app:SSE}
\begin{figure}[hbt]
    \centering
    \includegraphics[width=1\linewidth]{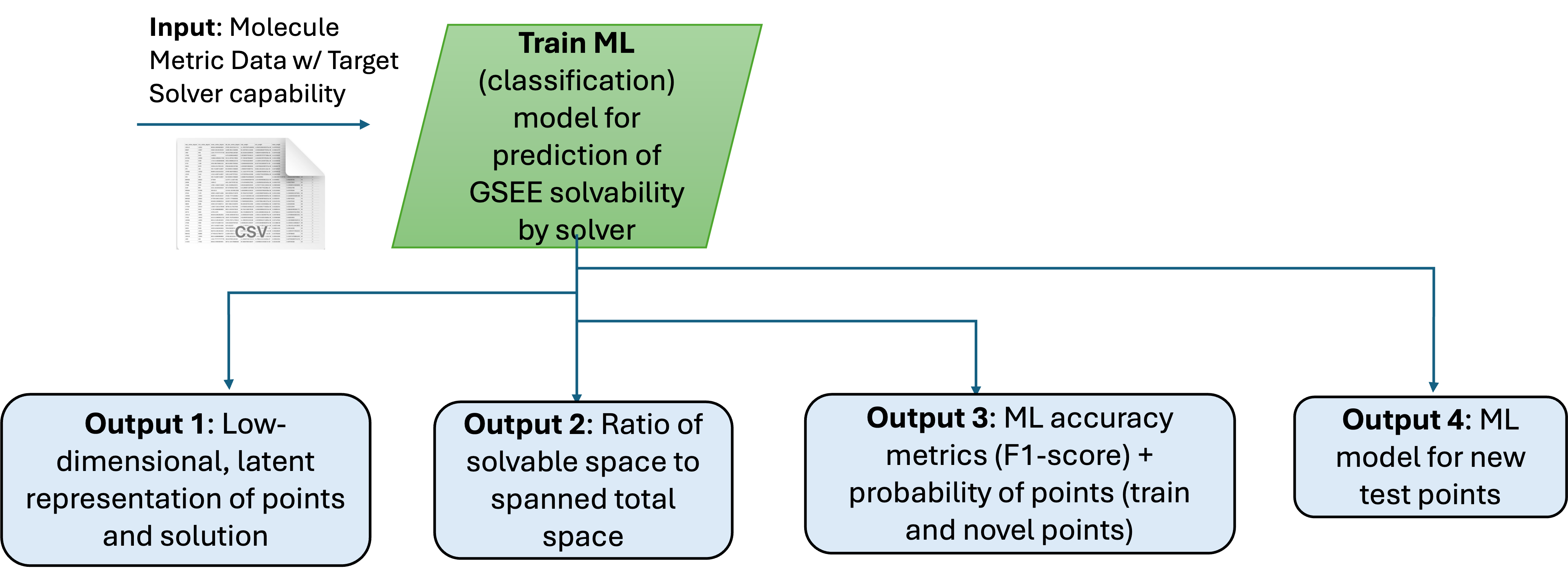}
    \caption{Machine Learning Framework for Ground State Energy Estimation Solvability Prediction}
    \label{fig:MLFramework}
\end{figure}

Given the problem features for a Hamiltonian and a performer's (or solver's) capability of solving it, we can use classic Machine Learning (ML) techniques for building a predictive model (which can predict a novel test point's solvability), and define a boundary between what would be classically solvable and what would not be, given the training data.  This allows us to estimate the ratio of the solvable space that the current dataset helps define.  Using the high-dimensional ML model learned using all the problem features, we compute an invertible latent space in which we can use the prediction probabilities from the ML model to define a ratio of solvable area to total area within the latent space.   While the original feature space could also be used to define this ratio, the sparsity of the original space (which is mitigated in the latent space) and the computational time are factors that led us to opt for computation in the latent space spanned by the problem features.  
% We provide more details of the method in
% Appendix \ref{app:ML}, 
We will provide more details of the method in
the appendix, 
but we outline the methodology here and provide an illustration of it in Figure \ref{fig:MLFramework}.

\begin{algorithm}[hbt!]

\caption{Methodology for the computation of solvable space}\label{alg:Solvability ratio}
\KwData{Data: $N$ rows of features in $R{^D}$ ($N \times D$ csv file); solver results ($N \times 1$ csv file), target\ results, confidence\ threshold, $thresh$}
\KwResult{Solvability ratio, SVM model, Classification metrics and visualization in latent space}
 $X \gets Data$\;
 $labels \gets target \ results$\;
 Fit a Support Vector Machine (SVM) model with $X, labels$  with hyper-parameter optimization and k-fold cross validation\;
Compute latent space with Non-negative Matrix Factorization, $X$$\approx WH$ to get lower-dimensional representation  $H$ in $R^{n}$ where ($n << D$)  and $W$, the transformed data in latent space $H$.  $X$ is first transformed with Min-max scaling.
Bound the latent space by computing min and max of $W$\;
Generate latent points, $W_{novel}$, between min and max for each latent dimension with resolution $r$\;
Inverse transform $W_{novel}$ points into original feature dimension to obtain $X_{novel}$ in $R{^D}$\;
Use SVM Model to predict the probability of every point in $X_{novel}$: $P(X | Solver = True)$\;
$Solvability\ Ratio$ = $\frac{|P(X_{novel} >= thresh|)}{|X_{novel}|}$ (where $\|...\| $ ) represents the cardinality or count within the set. \;
Compute classification metrics: Precision, Recall and F1-score, generated points along with the probability: $P(Target = True | Data)$ and latent visualization.\

\end{algorithm}

\begin{figure}[hbt!]
    \centering
    \includegraphics[width=1\linewidth]{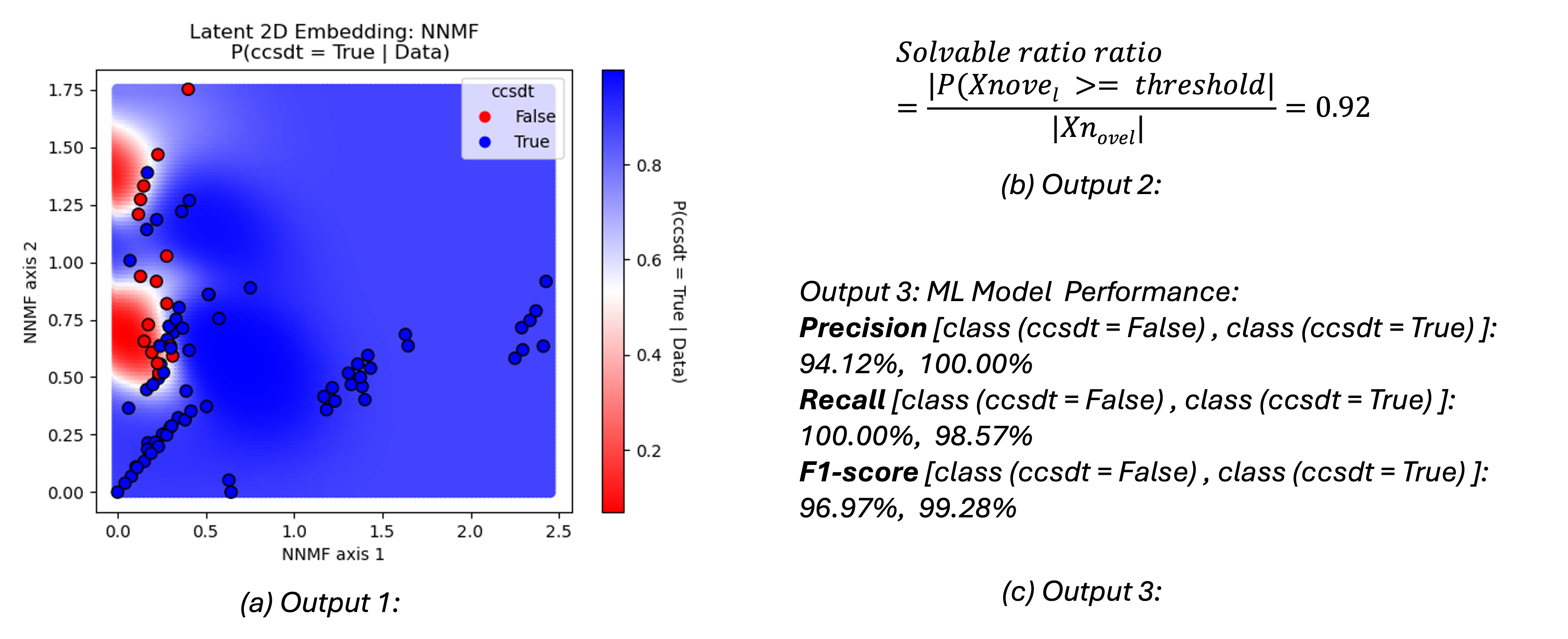}
    \caption{Example result with NNMF 2D Latent Space visualization with $P(ccsdt = True | Data)$ run with Data comprising of one norm of the Hamiltonian, log of fci dim,  number of electrons, number of orbitals and double factorized spectral gap.  Probabilities were computed from the high-dimensional SVM model but visualized in the 2D latent space.  Data points are also plotted in red (ccsdt = False) and blue (ccsdt = True)}
    \label{fig:ml_data_result}
\end{figure}

In our current dataset, we have 86 data points for the CCSDT solver.  However, presently,  out of the 86 rows, only 16 belong to the class of ccsdt = False.   Our goal is to eventually balance the dataset more to include more points for the negative class. Our solvability ratio for the current dataset will also reflect the imbalance since there is more information about the positive class, that is ccsdt = True.  An example of the outputs, which include solvability ratio, accuracy metrics, and latent space visualization, is shown in Figure \ref{fig:ml_data_result}.  This was run with features: one-norm of the Hamiltonian, log of the fci dimension,  number of electrons, number of orbitals, and the double factorized spectral gap.  Classification metrics are computed as defined in \cite{classification_metrics_implementation} and implemented by Python's scikit-learn library. 

\subsection{Importance Features} 
\begin{figure}[hbt!]
    \centering
    \includegraphics[width=1\linewidth]{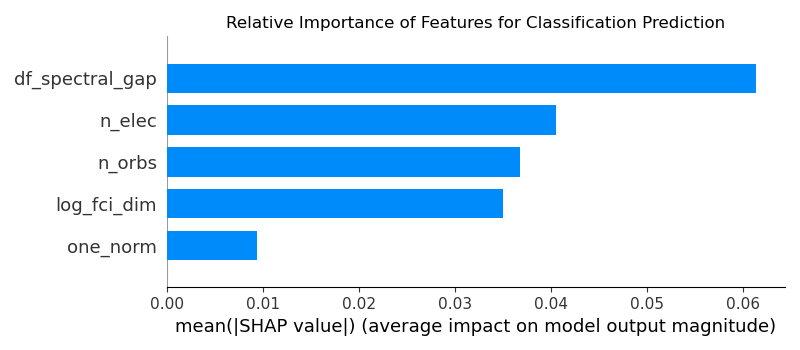}
    \caption{Relative Importance of features used for classification computed using the SHAP algorithm}
    \label{fig:ImportantFeatures_shap}
\end{figure}

We used Shapley Additive Explanations
\cite{shap} methodology, implemented by the Python library 
\cite{shap-library-repo}, to rank the relative importance features for the classification, as shown in Figure 
\ref{fig:ImportantFeatures_shap}.  
SHAP uses Shapley values from game theory to help feature importance values as well as other local explanations in the data.  
We see from Figure \ref{fig:ImportantFeatures_shap} that the feature double factorized spectral gap provides the most information as far as classification is concerned, followed by number of electrons, number of orbitals, and log of the fci dimension 
(all of relatively equal importance).  The least informative was the Hamiltonian one-norm.  

We have made a couple of choices in the framework for solvable space estimation that can affect the solvability ratio for the data.  We describe them below with justifications.

\subsection{Choice of Latent Space}
We chose the latent space computed by Non-Negative Matrix Factorization (NNMF) \cite{NNMF} as the dimensionality reduction algorithm because the dataset's Hamiltonian features in its current form consists of all positive values, therefore any inverse transformation from the latent space back into the original feature space should map to positive values as well.  Not all invertible dimensionality reduction methods guarantee this, but NNMF, by its inherent construction, does.  This choice was an important one because we use the latent space to generate new points (within its minimum- maximum span of its latent axes), and hence the generated, novel high-dimensional points pass the minimum criteria for physical validity (we do not yet have a filter for true physical validity for the generated points).  Another choice was to use auto-encoders to represent the latent space that we can generate new points from.  Auto-encoders are very powerful technologies that can represent the Hamiltonian features' latent representation because we simply do not have enough data to justify their use.  This may change in the future as users submit more data.   The total area could have also been defined within the span of the original feature space, without resorting to computing the latent space.  However, the original feature space is much more sparse compared to its latent space, and hence the computation time for the solvable area will take much longer.
 
\subsection{Choice of Machine Learning Model}
We chose the Support Vector Machine (SVM) \cite{svm} as our machine learning model.  In the past, we have tried out tree-based models (Random Forest \cite{RandomForest}, XGBoost \cite{xgboost}) as well, and the classification accuracy was similar to SVMs (given that all models were hyper-parameter optimized and cross-validated).  Given plenty of well-spanned data in the feature space, we could use tree-based or SVM models, but as far as having enough data is concerned, it is ideal to have 20-30 times the number of features with good coverage for all features to have robust results.  While one can use either of these models, we chose SVMs because the computed boundaries in the original feature space tend to be smoother than tree-based models, leading them to be mathematically described with fewer parameters.  Tree-based models tend to compute axis-aligned boundaries which require more parameters.   In the end, this point is not very important because, in most ML applications, one uses the trained model to make predictions on a test point and not really care about how complex the decision boundary is.  However,  in our case, it is possible that, in addition to using the trained model to make test predictions on new data, we would eventually want to mathematically describe our boundary in the original feature space.   Regardless of which model is eventually chosen to train the data on, it is imperative that it be run with hyper-parameter optimization and cross-validation.  In the accompanying user interface, we provide a choice of a Random Forest and an SVM model, along with choices of latent space computation methods so the user can look at the differences in the probability maps of the solution space in different latent spaces.

% \subsection{Active Learning} 
% Human learning in the real-world is an iterative process whereby we continually learn by interacting with the environment through feedback for our hypotheses. This allows us to change our theories or further confirm them.  Similarly, for our machine learning model that has an initial model for the boundary between classical solvability and intractability, we can have the model present to the experimenter those generated, novel Hamiltonian features whose solvability it is most uncertain about. The experimenter can run their solver on them, get an answer, and our model can be re-trained on the new data.  This allows us to refine our boundary in a statistically optimal way. 

% \section{Performance Metrics}\label{sec:performance_analysis}
% \input{025_performance_analysis}

\section{Benchmark Performer Examples}
% \label{sec:performer_examples}
% \input{301_solution_schema}
\subsection{Classical Performer Example - DMRG}
\label{sec:dmrg_example}

As another example of a classical performer, we include DMRG calculations. These calculations are performed similarly to those in Bellonzi et al. \cite{bellonzi2024}, with the addition of orbital ordering optimization. %({\color{magenta} DARPA QB Note: as per the updated, but not yet posted on Arxiv, version of Bellonzi et al.})
While we use additional computational time and extrapolation for any determination of reference energies by DMRG, we have elected for this performer example to only include the lowest variational energy obtained for a given set of calculations. This allows us to portray how such an approach may differ from the extrapolated approach. We also restrict the set of DMRG calculations of a single Hamiltonian (i.e. at different bond dimensions) to all run consecutively within 24hrs or to reflect the results as demonstrated in Bellonzi et al. \cite{bellonzi2024}. That is, we set the calculations to run starting at a bond dimension of 4 and the bond dimension  is automatically and repeatedly increased by 10\% until a convergence within $5\mathrm{x}10^{-5}$ Hartree is achieved or a 23.5 hour time limit is reached. There are times when the calculations converge to excited states. This means that the results do not necessarily reflect the best performance DMRG can achieve, but more what a solid first run may achieve. Any further refinements to the solutions would constitute a new solver and largely reflect how additional computational time and researcher time would improve the results.

This DMRG classical performer example has the solver ``short name'' \verb|DMRG_Niagara_cluster_lowest_energy| and the UUID \verb|16537433-9f4c-4eae-a65d-787dc3b35b59|, with the short name reflecting the method of energy selection and the use of the Niagara cluster. This cluster is hosted by SciNet \cite{ponce2019deploying,loken2010scinet}, who are  partnered with Compute Ontario and the Digital Research Alliance of Canada.

With the solver as defined above, we have addressed all 228 tasks available (as of the submission date) in the benchmark.

An example notebook that retrieves and solves a small subset of the problem instances with DMRG is included in the benchmark repository \cite{qb-gsee-benchmark-repo} as \verb|examples/run_dmrg.ipynb|. This notebook provides installation instructions for the benchmark and DMRG, parses the problem instance files, obtains the data files from the sftp server, runs DMRG and then produces solution files, validating them against the schema.

\subsection{Quantum Phase Estimation Resource Estimates}

% {\color{magenta} DARPA QB Note: Previous BOBQAT deliverables utilized different tools than those applied here for quantum resource estimation. Previously, OpenFermion was used for obtaining logical resource estimates and BenchQ for obtaining physical resource estimates through footprint analysis. Dec 10 and future deliverables instead utilize PyLIQTR for logical resource estimates. For physical resource estimates, the Dec 10 delivery uses a crude multiplicative factor to convert gate counts to runtimes as described below; we plan however to use a more defensible physical resource estimation methodology in future deliverables.}

As an example of a quantum resource estimation performer, we explore the Quantum Phase Estimation (QPE) algorithm.
These quantum resource estimates consider qubitized QPE with a double factorization of the electronic structure Hamiltonian \cite{vonBurg2021} as implemented by pyLIQTR \cite{pyliqtr}.
The failure and error analysis follows an approach similar to that of Bellonzi et al. \cite{bellonzi2024} and will be summarized here.
Algorithm parameters are chosen such that the ground-state energy will be estimated to within the target accuracy of 1.59~mHa with 99\% probability when accounting for imperfect ground-state overlap, hardware failure, and other error sources.
The initial state is taken to be the dominant Configuration State Function (CSF), whose ground-state overlap is estimated through the block2 \cite{zhai2023block2} implementation of DMRG.
Unlike the work of Bellonzi et al., the present work adopts a fixed truncation threshold for double-factorization of 1 mHa, rather than choosing the threshold based on the truncation error estimated by CCSD(T).
Furthermore, the error introduced by this truncation is assumed to be negligible.
This is motivated by the fact that prior studies have found that such a truncation threshold incurs errors that are comparable to chemical accuracy for large chemical systems, and also that the fact that demanding a smaller truncation error has relatively little impact on the logical resources required \cite{vonBurg2021,Lee2020}.

% \todo[inline,color=magenta]{TODO: Once we post an updated version of the catalysis paper to arxiv, we should update the text here and update the citation to specify v2.}

Several different physical costing models are used to estimate the runtime and physical qubit requirements.
These models use the Qualtran physical costing module \cite{qualtran} assuming four catalyzed AutoCCZ factories for magic state distillation \cite{Gidney2019efficientmagicstate} but make different assumptions about the hardware.
The optimistic hardware model of Lee et al. \cite{Lee2020} is used as a baseline.
This represents a \SI{1}{\micro\second} surface code cycle time and $10^{-4}$ physical error rate and corresponds to solver UUID \verb|5dad4064-cd11-412f-85cb-d722afe3b3de|.
A parallelized solver, corresponding to solver UUID \verb|4b07b89f-c66f-4e72-8c24-df3e4222cb41|, uses the same hardware model but assumes that enough QPUs are available to perform all shots in parallel.
In other words, its runtime for a given Hamiltonian is the estimated time to perform a single shot.
To explore the effects of hardware parameters, solver UUID \verb|2610d8de-bd3a-469e-9a80-473e8988755f| corresponds to a serial solver with the surface code cycle time reduced to \SI{0.1}{\micro\second}, solver UUID \verb|5d768520-b3d0-4292-bbb4-9776fa128107| corresponds to a serial solver with the physical error rate reduced to $10^{-6}$.
Lastly, solver UUID \verb|f6b36bde-be4a-4eee-975b-2c5f7e553f5f| corresponds to both reducing the surface code cycle time to \SI{0.1}{\micro\second} and the physical error rate to $10^{-6}$.
Note that these lowered surface code cycle times and physical error rates are well beyond the expected performance of superconducting qubits and are motivated by the possibility of future innovations.
Note also that for the benchmark problem instances, the architectures considered may not necessarily achieve the required hardware success rate, in which case no runtime is reported.

Note that there are a number of improvements that could significantly reduce the estimated runtime and qubit requirements.
This includes, for example, tensor hypercontraction block encodings \cite{Lee2020} and the incorporation of the Block-Invariant Symmetry Shift (BLISS) approach \cite{loaiza2023bliss,rocca2024scdf,caesura2025faster}, as well as using matrix product states to achieve a higher ground-state overlap \cite{fomichev2024stateprep,berry2024stateprep}.
Additionally, the true algorithm performance may be better than the performance model owing to the use of several upper bounds, as discussed in Ref. \cite{bellonzi2024}.

% \section{Appendix: Summary of Problem Instances}\label{app:prob_inst}
% \input{901_problem_instances.tex}

% \input{905_computational_status}

% \section{Appendix: Artifical Hamiltonian Generation}\label{app:art_hamil}
% \input{902_artifical_hamiltonians}

\printbibliography

\end{document}